\documentclass[aps,prd,superscriptaddress]{revtex4}
\usepackage{epsfig,epsf}
\usepackage{amsmath}
\usepackage{amsthm}
\usepackage{amsfonts}
\usepackage{amssymb}
\usepackage{dsfont}
\usepackage{multirow}
\usepackage{appendix}
\usepackage{slashed}
\usepackage[active]{srcltx}
\usepackage{psfrag}

\begin{document}

\title{Hadronic molecules $\eta_c \eta_c$ and $\chi_{c0}\chi_{c0}$}
\date{\today}
\author{S.~S.~Agaev}
\affiliation{Institute for Physical Problems, Baku State University, Az--1148 Baku,
Azerbaijan}
\author{K.~Azizi}
\thanks{kazem.azizi@ut.ac.ir, Corresponding author}
\affiliation{Department of Physics, University of Tehran, North Karegar Avenue, Tehran
14395-547, Iran}
\affiliation{Department of Physics, Do\v{g}u\c{s} University, Dudullu-\"{U}mraniye, 34775
Istanbul, T\"{u}rkiye}
\author{B.~Barsbay}
\affiliation{Division of Optometry, School of Medical Services and Techniques, Do\v{g}u%
\c{s} University, 34775 Istanbul, T\"{u}rkiye}
\author{H.~Sundu}
\affiliation{Department of Physics Engineering, Istanbul Medeniyet University, 34700
Istanbul, T\"{u}rkiye}

\begin{abstract}
The fully charmed hadronic scalar molecules $\mathcal{M}_1=\eta_c \eta_c$
and $\mathcal{M}_2=\chi_{c0}\chi_{c0}$ are studied in the context of the QCD
sum rule method. The masses $m$, $\widetilde{m}$ and current couplings $f$, $%
\widetilde{f}$ of these states are calculated using the two-point sum rule
approach. The obtained results $m=(6264 \pm 50)~\mathrm{MeV}$ and $%
\widetilde{m}=(6954 \pm 50)~\mathrm{MeV}$ are employed to determine their
decay channels. It is demonstrated that the processes $\mathcal{M}_1\to
J/\psi J/\psi $ and $\mathcal{M}_1\to \eta _{c}\eta _{c}$ are kinematically
allowed decay modes of $\mathcal{M}_1$. The molecule $\mathcal{M}_2$ decays
to $J/\psi J/\psi$, $J/\psi \psi^{\prime}$, $\eta _{c}\eta _{c}$, $\eta
_{c}\eta _{c}(2S)$, $\eta _{c}\chi _{c1}(1P)$, and $\chi_{c0} \chi_{c0}$
mesons. The partial widths all of these processes are evaluated by means of
the three-point sum rule calculations, which are necessary to extract the
strong couplings $g_i$ at vertices $\mathcal{M}_1J/\psi J/\psi $, $\mathcal{M%
}_1\eta _{c}\eta _{c}$, and others. Our estimates for the full widths of the
molecules $\Gamma_{\mathcal{M}_1}=(320 \pm 72)~\mathrm{MeV}$ and $\Gamma _{%
\mathcal{M}_2}=(138 \pm 18)~\mathrm{MeV}$, as well as their masses are
compared with parameters of the $X$ resonances discovered by the
LHCb-ATLAS-CMS Collaborations in the di-$J/\psi$ and $J/\psi\psi^{\prime}$
invariant mass distributions. We argue that the molecule $\mathcal{M}_1$ can
be considered as a real candidate to the resonance $X(6200)$. The structure $%
\mathcal{M}_2$ may be interpreted as $X(6900)$ or one of its components in
combination with a scalar tetraquark.
\end{abstract}

\maketitle


\section{Introduction}

\label{sec:Int} 

The discovery of the resonances $X(6200)$, $X(6600)$, $X(6900)$, and $%
X(7300) $ in the di-$J/\psi $ and $J/\psi \psi ^{\prime }$ invariant mass
distributions by the LHCb, ATLAS, and CMS Collaborations gave new impetus to
investigations of fully charmed and beauty four-quark mesons \cite%
{LHCb:2020bwg,Bouhova-Thacker:2022vnt,CMS:2023owd}. Heavy exotic mesons
composed of two or four $c$ and $b$ quarks attracted interest of researches
already at early stages of multiquark hadrons' physics \cite%
{Iwasaki:1975pv,Chao:1980dv,Ader:1981db,Lipkin:1986dw,Zouzou:1986qh,
Heller:1985cb,Carlson:1987hh,Barnea:2006sd,Vijande:2007ix,Ebert:2007rn}.
These investigations were continued in more recent papers \cite%
{Berezhnoy:2011xn,Chen:2016jxd,
Wu:2016vtq,Bai:2016int,Wang:2017jtz,Debastiani:2017msn,Richard:2017vry,Anwar:2017toa, Richard:2018yrm,Esposito:2018cwh,Liu:2019zuc,Bedolla:2019zwg}%
.

One of main problems studied in pioneering articles was stability of such
hadrons against strong decays. It was argued that tetraquarks containing a
heavy diquark and a light antidiquark may be strong-interaction stable
particles, whereas fully heavy structures are unstable against strong decays
(see, for instance, Refs.\ \cite%
{Ader:1981db,Lipkin:1986dw,Zouzou:1986qh,Carlson:1987hh}). Detailed
quantitative explorations led to different conclusions concerning allowed
decay channels of fully charmed or beauty exotic mesons. Thus, in accordance
with Ref.\ \cite{Berezhnoy:2011xn}, the scalar and axial-vector tetraquarks $%
X_{\mathrm{4c}}=cc\overline{c}\overline{c}$ cannot decay to $J/\psi J/\psi $
mesons, because their masses are less than the di-$J/\psi $ threshold. Only
a mass of a tensor tetraquark $X_{\mathrm{4c}}$ exceeds this limit and can
be seen in the di-$J/\psi $ mass distribution. At the same time, all fully
beauty structures $X_{\mathrm{4b}}$ are below $\Upsilon (1S)\Upsilon (1S)$
threshold, therefore do not transform strongly to these mesons. Tetraquarks $%
X_{\mathrm{4c}}$ and $X_{\mathrm{4b}}$ with different spin-parities were
studied in Ref.\ \cite{Chen:2016jxd}, in which it was demonstrated that
scalar $X_{\mathrm{4c}}$ decays to $\eta _{c}\eta _{c}$, $J/\psi J/\psi $,
and $\eta _{c}\chi _{c1}(1P)$ mesons, whereas $X_{\mathrm{4b}}$ is stable
against strong decays to two bottomonia expect for a scalar
diquark-antidiquark state $X_{\mathrm{4b}}$ built of pseudoscalar components.

Information of the LHCb Collaboration generated new publications aimed to
explain origin of observed new structures, calculate their masses and
explore possible decay channels \cite%
{Zhang:2020xtb,Albuquerque:2020hio,Becchi:2020mjz,Becchi:2020uvq,Dong:2020nwy,Liang:2021fzr}%
. Thus, the mass of the scalar tetraquark\ $X_{\mathrm{4c}}$ was estimated
around $6.44-6.47~\mathrm{GeV}$ in the framework of QCD sum rule method \cite%
{Zhang:2020xtb}, and the author interpreted $X_{\mathrm{4c}}$ as a part of a
threshold enhancement $6.2-6.8\ \mathrm{GeV}$ seen by LHCb in nonresonant di-%
$J/\psi $ production. The hadronic molecule $\chi _{c0}\chi _{c0}$ or/and
the diquark-antidiquark state with pseudoscalar constituents were considered
as candidates to the resonance $X(6900)$ in Ref.\ \cite{Albuquerque:2020hio}%
. Decay channels of the fully heavy tetraquarks to conventional leptons and
mesons through $Q\overline{Q}$ annihilations were investigated in Refs.\
\cite{Becchi:2020mjz,Becchi:2020uvq}.

The LHCb data were analyzed in Ref.\ \cite{Dong:2020nwy} in the context of a
coupled-channel method: It was argued that in the di-$J/\psi $ system there
is a near-threshold state $X(6200)$ having the spin-parities $0^{++}$ or $%
2^{++}$. The coupled-channel effects may also produce a pole structure,
which was denoted as the resonance $X(6900)$ in Ref.\ \cite{Liang:2021fzr}.
The performed analysis helped the authors to declare also the existence of a
bound state $X(6200)$, and the broad and narrow resonances $X(6680)$ and $%
X(7200)$, respectively.

The discoveries of the ATLAS and CMS experiments intensified analyses of new
heavy $X$ resonances \cite%
{Wang:2022xja,Dong:2022sef,Faustov:2022mvs,Niu:2022vqp,Yu:2022lak,Kuang:2023vac}%
. In fact, in Ref.\ \cite{Wang:2022xja} the $X(6200)$ was considered to be
the ground-level tetraquark structure with $J^{\mathrm{PC}}=0^{++}$ or $%
1^{+-}$, whereas its first radial excitation was assigned as $X(6600)$.
Similar interpretations were extended to the whole family of heavy $X$
structures in Ref.\ \cite{Dong:2022sef}, where the authors suggested to
consider the resonances $X(6200)-X(7300)$ as $1S$, $1P/2S$, $1D/2P$, and $%
2D/3P/4S$ tetraquark states. Close ideas were proposed in the context of the
relativistic quark model as well \cite{Faustov:2022mvs}.

It is clear, that a wide variety of alternatives to explain the experimental
data makes important detailed investigations of fully heavy tetraquarks. In
our article \cite{Agaev:2023wua}, we calculated the masses of the scalar
diquark-antidiquark states $X_{\mathrm{4c}}$ and $X_{\mathrm{4b}}$ built of
axial-vector constituents, and estimated the full width of $X_{\mathrm{4c}}$%
. Our results for the mass $m=(6570\pm 55)~\mathrm{MeV}$ and width $\Gamma _{%
\mathrm{4c}}=(110\pm 21)~\mathrm{MeV}$ of the tetraquark $X_{\mathrm{4c}}$
allowed us to consider it as a candidate to the resonance $X(6600)$. Relying
on their decay channels $X(6600)\rightarrow J/\psi J/\psi $ and $%
X(7300)\rightarrow J/\psi \psi ^{\prime }$, we also supposed that $X(7300)$
may be $2S$ excitation of $X(6600)$: Here, we took into account that $\psi
^{\prime }$~is $2S$ excited state of the meson $J/\psi $. We computed the
mass of the fully beauty scalar state $X_{\mathrm{4b}}$ and got $m^{\prime
}=(18540\pm 50)~\mathrm{MeV}$ which is below the $\eta _{b}\eta _{b}$
threshold. Hence $X_{\mathrm{4b}}$ cannot decay to hidden-bottom mesons,
i.e., this tetraquark is observable neither in the $\eta _{b}\eta _{b}$ nor
in $\Upsilon (1S)\Upsilon (1S)$ mass distributions. The break-up of $X_{%
\mathrm{4b}}$ to ordinary mesons proceeds through its strong decays to
open-bottom mesons, or via electroweak leptonic and nonleptonic processes.

The scalar diquark-antidiquark states $T_{\mathrm{4c}}$ and $T_{\mathrm{4b}}$
with pseudoscalar components were explored in Ref.\ \cite{Agaev:2023gaq}, in
which we computed spectroscopic parameters of these tetraquarks. We
interpreted the tetraquark $T_{\mathrm{4c}}$ with the mass $m=(6928\pm 50)~%
\mathrm{MeV}$ and width $\widetilde{\Gamma }_{\mathrm{4c}}=(128\pm 22)~%
\mathrm{MeV}$ as a resonance $X(6900)$. The mass and width of its beauty
counterpart $T_{\mathrm{4b}}$ were found equal to $m^{\prime }=(18858\pm 50)~%
\mathrm{MeV}$ and $\widetilde{\Gamma }_{\mathrm{4b}}=(94\pm 28)~\mathrm{MeV}$%
, respectively.

In present article, we explore the hadronic molecules $\mathcal{M}_{1}=\eta
_{c}\eta _{c}$ and $\mathcal{M}_{2}=\chi _{c0}\chi _{c0}$ by computing their
masses and widths to confront obtained predictions with the both available
experintal data and results of diquark-antidiquark model. The masses of
these structures are evaluated using the QCD two-point sum rule method. To
estimate their widths, we apply the three-point sum rule approach, which is
required to extract strong couplings $g_{i}$ at vertices, for example, $%
\mathcal{M}_{1}J/\psi J/\psi $ and $\mathcal{M}_{1}\eta _{c}\eta _{c}$ in
the case of $\mathcal{M}_{1}$.

This paper is structures in the following way: In Section \ref{sec:MW1}, we
calculate the mass and current coupling of the molecule $\mathcal{M}_{1}$.
We evaluate its full width using strong processes $\mathcal{M}%
_{1}\rightarrow J/\psi J/\psi $ and $\mathcal{M}_{1}\rightarrow \eta
_{c}\eta _{c}$. \ In Section \ref{sec:MW2}, we analyze in a detailed form
spectroscopic parameters of the molecule $\mathcal{M}_{2}$. The full width
of $\mathcal{M}_{2}$ is found by considering decays $\mathcal{M}%
_{2}\rightarrow J/\psi J/\psi $, $J/\psi \psi ^{\prime }$, $\eta _{c}\eta
_{c}$, $\eta _{c}\eta _{c}(2S)$, $\eta _{c}\chi _{c1}(1P)$ and $\mathcal{M}%
_{2}\rightarrow \chi _{c0}\chi _{c0}$. Last section is reserved for
discussion of results and concluding remarks. Appendix contains the explicit
expression of the heavy-quark propagator, and different correlation
functions employed in the analyses.


\section{Mass, current coupling and width of the molecule $\protect\eta _{c}%
\protect\eta _{c}$}

\label{sec:MW1} 

In this section, we compute the mass $m$, current coupling $f$ and full
width $\Gamma _{\mathcal{M}_{1}}$ of the hadronic molecule $\mathcal{M}%
_{1}=\eta _{c}\eta _{c}$ using the QCD sum rule method \cite%
{Shifman:1978bx,Shifman:1978by}.


\subsection{Mass and coupling}

To derive the two-point sum rules for the mass $m$ and current coupling $f$
of the molecule $\mathcal{M}_{1}$, we explore the two-point correlation
function%
\begin{equation}
\Pi (p)=i\int d^{4}xe^{ipx}\langle 0|\mathcal{T}\{J(x)J^{\dag
}(0)\}|0\rangle ,  \label{eq:CF1}
\end{equation}%
where, $\mathcal{T}$ is the time-ordered product of two currents, and $J(x)$
is the interpolating current for the molecule $\mathcal{M}_{1}$.

The current for $\mathcal{M}_{1}$ reads
\begin{equation}
J(x)=\overline{c}_{a}(x)i\gamma _{5}c_{a}(x)\overline{c}_{b}(x)i\gamma
_{5}c_{b}(x),  \label{eq:CR1}
\end{equation}%
where $a$, and $b$ are color indices. It describes a hadronic molecule with
spin-parities $J^{\mathrm{PC}}=0^{++}$.

The physical side of the sum rule $\Pi ^{\mathrm{Phys}}(p)$ can be obtained
from Eq.\ (\ref{eq:CF1}) by inserting a complete set of intermediate states
with quark content and spin-parities of the molecule $\mathcal{M}_{1}$, and
carrying out integration over $x$
\begin{equation}
\Pi ^{\mathrm{Phys}}(p)=\frac{\langle 0|J|\mathcal{M}_{1}(p)\rangle \langle
\mathcal{M}_{1}(p)|J^{\dagger }|0\rangle }{m^{2}-p^{2}}+\cdots .
\label{eq:Phys1}
\end{equation}%
In Eq.\ (\ref{eq:Phys1}) the ground-state contribution is presented
explicitly, whereas higher resonances and continuum terms are denoted by the
ellipses.

The function $\Pi ^{\mathrm{Phys}}(p)$ can be rewritten using the matrix
element of the molecule $\mathcal{M}_{1}$
\begin{equation}
\langle 0|J|\mathcal{M}_{1}(p)\rangle =fm,  \label{eq:ME1}
\end{equation}%
which leads to the following expression
\begin{equation}
\Pi ^{\mathrm{Phys}}(p)=\frac{f^{2}m^{2}}{m^{2}-p^{2}}+\cdots .
\label{eq:Phen2}
\end{equation}%
The correlator $\Pi ^{\mathrm{Phys}}(p)$ has a Lorentz structure which is
proportional to $\mathrm{I}$. Consequently, corresponding invariant
amplitude $\Pi ^{\mathrm{Phys}}(p^{2})$ is equal to the expression in
right-hand side of Eq.\ (\ref{eq:Phen2}).

The second component of the sum rule analysis, i.e., the function $\Pi ^{%
\mathrm{OPE}}(p)$ should be calculated in the operator product expansion ($%
\mathrm{OPE}$) with some accuracy. In terms of the $c$-quark propagators $%
S_{c}(x)$ the function $\Pi ^{\mathrm{OPE}}(p)$ has the following form
\begin{eqnarray}
&&\Pi ^{\mathrm{OPE}}(p)=i\int d^{4}xe^{ipx}\left\{ \mathrm{Tr}\left[ \gamma
_{5}S_{c}^{ba^{\prime }}(x)\gamma _{5}S_{c}^{a^{\prime }b}(-x)\right]
\mathrm{Tr}\left[ \gamma _{5}S_{c}^{ab^{\prime }}(x)\gamma
_{5}S_{c}^{b^{\prime }a}(-x)\right] \right.  \notag \\
&&-\mathrm{Tr}\left[ \gamma _{5}S_{c}^{bb^{\prime }}(x)\gamma
_{5}S_{c}^{b^{\prime }a}(-x)\gamma _{5}S_{c}^{aa^{\prime }}(x)\gamma
_{5}S_{c}^{a^{\prime }b}(-x)\right] -\mathrm{Tr}\left[ \gamma
_{5}S_{c}^{ba^{\prime }}(x)\gamma _{5}S_{c}^{a^{\prime }a}(-x)\gamma
_{5}\right.  \notag \\
&&\left. \left. \times S_{c}^{ab^{\prime }}(x)\gamma _{5}S_{c}^{b^{\prime
}b}(-x)\right] +\mathrm{Tr}\left[ \gamma _{5}S_{c}^{bb^{\prime }}(x)\gamma
_{5}S_{c}^{b^{\prime }b}(-x)\right] \mathrm{Tr}\left[ \gamma
_{5}S_{c}^{aa^{\prime }}(x)\gamma _{5}S_{c}^{a^{\prime }a}(-x)\right]
\right\} .  \label{eq:QCD1}
\end{eqnarray}%
The propagator $S_{c}(x)$ contains terms which are linear and quadratic in
gluon field strength. As a result, $\Pi ^{\mathrm{OPE}}(p)$ does not depend
on light quark or mixed quark-gluon vacuum condensates. The explicit
expression of $S_{c}(x)$ can be found in Appendix (see, also Ref.\ \cite%
{Agaev:2020zad}).

The $\Pi ^{\mathrm{OPE}}(p)$ has also a Lorentz structure proportional $%
\mathrm{I}$. We denote the corresponding invariant amplitude as $\Pi ^{%
\mathrm{OPE}}(p^{2})$. To find a sum rule equality one has to equate the
functions $\Pi ^{\mathrm{Phys}}(p^{2})$ and $\Pi ^{\mathrm{OPE}}(p^{2})$,
apply the Borel transformation for suppressing contributions of higher
resonances and continuum states, and subtract suppressed terms using the
assumption about quark-hadron duality \cite{Shifman:1978bx,Shifman:1978by}.
After these manipulations the amplitude $\Pi ^{\mathrm{OPE}}(p^{2})$ becomes
a function of the Borel and continuum subtraction parameters $M^{2}$ and $%
s_{0}$, and will be denoted by $\Pi (M^{2},s_{0})$.

Calculation of $\Pi (M^{2},s_{0})$ is a next step to derive the sum rules
for the mass $m$ and coupling $f$. Analyses demonstrate that $\Pi
(M^{2},s_{0})$ has the form%
\begin{equation}
\Pi (M^{2},s_{0})=\int_{16m_{c}^{2}}^{s_{0}}ds\rho ^{\mathrm{OPE}%
}(s)e^{-s/M^{2}}.  \label{eq:InvAmp}
\end{equation}%
where $\rho ^{\mathrm{OPE}}(s)$ is a two-point spectral density found as an
imaginary part of the invariant amplitude $\Pi ^{\mathrm{OPE}}(p^{2})$. The
function $\rho ^{\mathrm{OPE}}(s)$ contains a perturbative term $\rho ^{%
\mathrm{pert.}}(s)$ and a dimension-$4$ nonperturbative contribution $\sim
\langle \alpha _{s}G^{2}/\pi \rangle $. An analytical expression of $\rho ^{%
\mathrm{OPE}}(s)$ is rather cumbersome, therefore we do not write it here
explicitly.

The mass $m$ and coupling $f$ can be extracted from the sum rules
\begin{equation}
m^{2}=\frac{\Pi ^{\prime }(M^{2},s_{0})}{\Pi (M^{2},s_{0})}  \label{eq:Mass}
\end{equation}%
and
\begin{equation}
f^{2}=\frac{e^{m^{2}/M^{2}}}{m^{2}}\Pi (M^{2},s_{0}),  \label{eq:Coupl}
\end{equation}%
respectively. In Eq.\ (\ref{eq:Mass}), we introduce a notation $\Pi ^{\prime
}(M^{2},s_{0})=d\Pi (M^{2},s_{0})/d(-1/M^{2})$.

The parameters which enter to these sum rules are the gluon vacuum
condensate $\langle \alpha _{s}G^{2}/\pi \rangle $ and the mass of $c$
quark. Their numerical values are presented below%
\begin{equation}
\langle \frac{\alpha _{s}G^{2}}{\pi }\rangle =(0.012\pm 0.004)~\mathrm{GeV}%
^{4},\ m_{c}=(1.27\pm 0.02)~\mathrm{GeV}.  \label{eq:Parameters}
\end{equation}

A choice of the working regions for $M^{2}$ and $s_{0}$ is another problem
of sum rule analyses. These parameters should be determined in such a way
that to satisfy the requirement imposed by the pole contribution ($\mathrm{PC%
}$), and ensure convergence of the operator product expansion. Prevalence of
a perturbative contribution over a nonperturbative one, as well as stability
of physical quantities under variation of these parameters are also among
important constraints.

Because, in the present article we consider a nonperturbative term $\sim
\langle \alpha _{s}G^{2}/\pi \rangle $, the pole contribution plays a key
role in determination of the working intervals for the $M^{2}$ and $s_{0}$.
To estimate $\mathrm{PC}$, we use the formula%
\begin{equation}
\mathrm{PC}=\frac{\Pi (M^{2},s_{0})}{\Pi (M^{2},\infty )},  \label{eq:PC}
\end{equation}%
and require fulfillment of the constraint $\mathrm{PC}\geq 0.5$.

\begin{figure}[h]
\includegraphics[width=8.5cm]{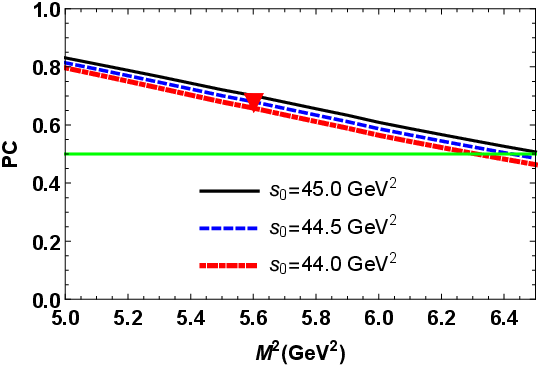}
\caption{The pole contribution $\mathrm{PC}$ as a function of the Borel
parameter $M^{2}$ at different $s_{0}$. The limit $\mathrm{PC}=0.5$ is
plotted by the horizontal line. The red triangle shows the point, where the
mass $m$ of $\mathcal{M}_{1}$ has effectively been extracted from the sum
rule. }
\label{fig:PC}
\end{figure}
The $\mathrm{PC}$ is employed to fix the higher limit of the Borel parameter
$M^{2}$. The lower limit for $M^{2}$, in the case under discussion, is found
from a stability of the sum rules' results under variation of $M^{2}$, and
from superiority of the perturbative term. Two values of $M^{2}$ extracted
by this way fix boundaries of the region where $M^{2}$ can be varied.
\qquad\

Calculations for the molecule $\mathcal{M}_{1}$ prove that intervals
\begin{equation}
M^{2}\in \lbrack 5,6.5]~\mathrm{GeV}^{2},\ s_{0}\in \lbrack 44,45]~\mathrm{%
GeV}^{2},  \label{eq:Wind1}
\end{equation}%
are suitable regions for the parameters $M^{2}$ and $s_{0}$, where they
comply with limits on $\mathrm{PC}$ and nonperturbative term. Thus, at $%
M^{2}=6.5~\mathrm{GeV}^{2}$ the pole contribution is $0.49$, whereas at $%
M^{2}=5~\mathrm{GeV}^{2}$ it becomes equal to $0.81$. At the minimum of $%
M^{2}=5~\mathrm{GeV}^{2}$, contribution of the nonperturbative term forms $%
\simeq 5\%$ of the correlation function. In Fig.\ \ref{fig:PC}, we plot $%
\mathrm{PC}$ as a function of $M^{2}$ at different $s_{0}$ to show its
changes in explored range of $\ M^{2}$. It is clear, that the pole
contribution overshoots $0.5$ for all values of the parameters $M^{2}$ and $%
s_{0}$ from Eq.\ (\ref{eq:Wind1}) excluding very small region around of the
point $M^{2}=6.5~\mathrm{GeV}^{2}$.

\begin{figure}[h!]
\begin{center}
\includegraphics[totalheight=6cm,width=8cm]{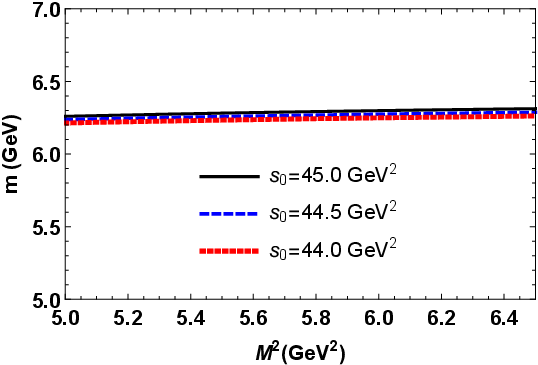}\,\, %
\includegraphics[totalheight=6cm,width=8cm]{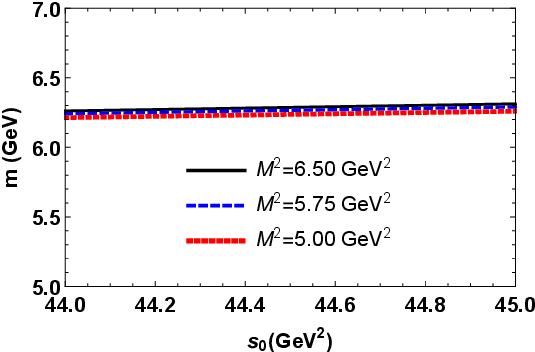}
\end{center}
\caption{ Mass of the hadronic molecule $\mathcal{M}_1$ as a function of the
Borel $M^2$ (left), and the continuum threshold $s_0$ parameters (right).}
\label{fig:Mass}
\end{figure}

The mass $m$ and coupling $f$ of the molecule $\mathcal{M}_{1}$ are
determined by calculating them at different $M^{2}$ and $s_{0}$ from the
regions Eq.\ (\ref{eq:Wind1}), and averaging obtained results to find mean
values of these parameters. Final results for $m$ and $f$ are
\begin{equation}
m=(6264\pm 50)~\mathrm{MeV},\ \ f=(2.12\pm 0.16)\times 10^{-2}~\mathrm{GeV}%
^{4}.  \label{eq:Result1}
\end{equation}%
The predictions Eq.\ (\ref{eq:Result1}) correspond to sum rules' results at
the point $M^{2}=5.6~\mathrm{GeV}^{2}$ and $s_{0}=44.5~\mathrm{GeV}^{2}$
which is approximately at a middle of the regions Eq.\ (\ref{eq:Wind1}). At
this point the pole contribution is $\mathrm{PC}\approx 0.68$ which ensures
superiority of $\mathrm{PC}$ in the results, and confirms ground-level
nature of the molecule $\mathcal{M}_{1}$. Dependence of $m$ on the
parameters $M^{2}$ and $s_{0}$ is plotted in Fig.\ \ref{fig:Mass}.

Our result for $m$ nicely agrees with the mass of the resonance $X(6200)$

\begin{equation}
m^{\mathrm{ATL}}=6220\pm 50_{-50}^{+40}~\mathrm{MeV},  \label{eqMWATL1}
\end{equation}%
reported by the ATLAS Collaboration \cite{Bouhova-Thacker:2022vnt}. But for
reliable conclusions on a nature of the resonance $X(6200)$ there is a
necessity also to estimate the full width of the molecule $\mathcal{M}_{1}$:
Below, we provide results of relevant studies.

\subsection{Full width}


The mass of the hadronic molecule $\mathcal{M}_{1}$ exceeds the two-meson $%
J/\psi J/\psi $ and $\eta _{c}\eta _{c}$ thresholds $6192~\mathrm{MeV}$ and $%
5968~\mathrm{MeV}$, respectively. Hence $S$-wave decay channels $\mathcal{M}%
_{1}\rightarrow J/\psi J/\psi $ and $\mathcal{M}_{1}\rightarrow \eta
_{c}\eta _{c}$ are allowed modes of this particle.


\subsubsection{Decay $\mathcal{M}_{1}\rightarrow J/\protect\psi J/\protect%
\psi $}


We start our studies from consideration of the decay $\mathcal{M}%
_{1}\rightarrow J/\psi J/\psi $. Partial width of this process is determined
by the strong coupling $g_{1}$ at the vertex $\mathcal{M}_{1}J/\psi J/\psi $%
. In the framework of the QCD sum rule method $g_{1}$ can be obtained from
analysis of the three-point correlation function%
\begin{equation}
\Pi _{\mu \nu }(p,p^{\prime })=i^{2}\int d^{4}xd^{4}ye^{ip^{\prime
}y}e^{-ipx}\langle 0|\mathcal{T}\{J_{\mu }^{\psi }(y)J_{\nu }^{\psi
}(0)J^{\dagger }(x)\}|0\rangle ,  \label{eq:CF2}
\end{equation}%
where $J_{\mu }^{\psi }(x)$ is the interpolating currents for the vector
meson $J/\psi $
\begin{equation}
J_{\mu }^{\psi }(x)=\overline{c}_{i}(x)\gamma _{\mu }c_{i}(x),
\label{eq:CR2}
\end{equation}%
where $i=1,2,3$ are the color indices. The $4$-momentum of the molecule $%
\mathcal{M}_{1}$ is $p$, whereas momenta of the $J/\psi $ mesons are $%
p^{\prime }$ and $q=p-p^{\prime }$, respectively.

After some calculations, for the physical side of the sum rule, we find
\begin{eqnarray}
&&\Pi _{\mu \nu }^{\mathrm{Phys}}(p,p^{\prime })=g_{1}(q^{2})\frac{%
fmf_{1}^{2}m_{1}^{2}}{\left( p^{2}-m^{2}\right) \left( p^{\prime
2}-m_{1}^{2}\right) (q^{2}-m_{1}^{2})}  \notag \\
&&\times \left[ \frac{1}{2}\left( m^{2}-m_{1}^{2}-q^{2}\right) g_{\mu \nu
}-q_{\mu }p_{\nu }^{\prime }\right] +\cdots ,  \label{eq:CR2A}
\end{eqnarray}%
where $m_{1}$ and $f_{1}$ are the mass and decay constant of the $J/\psi $
meson. To derive Eq.\ (\ref{eq:CR2A}), we have isolated the contribution of
the ground-state particles from other terms, and made use the following
matrix elements
\begin{equation}
\langle 0|J_{\mu }^{\psi }|J/\psi (p)\rangle =f_{1}m_{1}\varepsilon _{\mu
}(p),  \label{eq:ME2}
\end{equation}%
and

\begin{equation}
\langle J/\psi (p^{\prime })J/\psi (q)|\mathcal{M}_{1}(p)\rangle
=g_{1}(q^{2})\left[ q\cdot p^{\prime }\varepsilon ^{\ast }(p^{\prime })\cdot
\varepsilon ^{\ast }(q)-q\cdot \varepsilon ^{\ast }(p^{\prime })p^{\prime
}\cdot \varepsilon ^{\ast }(q)\right] .  \label{eq:ME3}
\end{equation}

The correlator $\Pi _{\mu \nu }^{\mathrm{Phys}}(p,p^{\prime })$ contains two
Lorentz structures that can be used to obtain the sum rule for $g_{1}(q^{2})$%
. We select to work with the term $\sim g_{\mu \nu }$ and present the
relevant invariant amplitude by $\Pi ^{\mathrm{Phys}}(p^{2},p^{\prime
2},q^{2})$. The Borel transformations over $p^{2}$ and $p^{\prime 2}$ of the
amplitude $\Pi ^{\mathrm{Phys}}(p^{2},p^{\prime 2},q^{2})$ yield
\begin{equation}
\mathcal{B}\Pi ^{\mathrm{Phys}}(p^{2},p^{\prime
2},q^{2})=g_{1}(q^{2})fmf_{1}^{2}m_{1}^{2}\frac{m^{2}-m_{1}^{2}-q^{2}}{%
2(q^{2}-m_{1}^{2})}e^{-m^{2}/M_{1}^{2}}e^{-m_{1}^{2}/M_{2}^{2}}+\cdots .
\label{eq:CorrF5a}
\end{equation}

The correlation function $\Pi _{\mu \nu }(p,p^{\prime })$ calculated in
terms of $c$-quark propagators reads
\begin{eqnarray}
&&\Pi _{\mu \nu }^{\mathrm{OPE}}(p,p^{\prime })=-2\int
d^{4}xd^{4}ye^{ip^{\prime }y}e^{-ipx}\mathrm{Tr}\left[ \gamma _{\mu
}S_{c}^{ib}(y-x)\right.  \notag \\
&&\left. \times \gamma _{5}S_{c}^{bj}(x){}\gamma _{\nu }S_{c}^{ja}(-x)\gamma
_{5}S_{c}^{ai}(x-y)\right].  \label{eq:QCDside}
\end{eqnarray}

\begin{figure}[h]
\includegraphics[width=8.5cm]{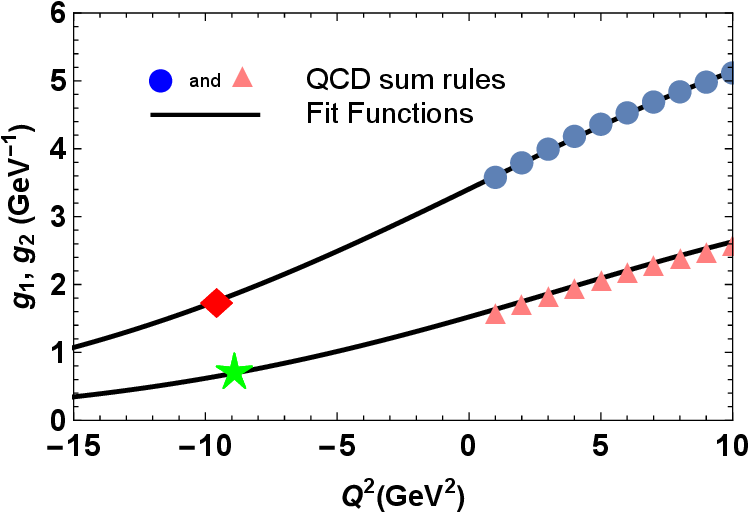}
\caption{The sum rule predictions and fit functions for the strong couplings
$g_{1}(Q^{2})$ (upper line) and $g_{2}(Q^{2})$ (lower line). The red diamond
and green star denote the points $Q^{2}=-m_{1}^{2}$ and $Q^{2}=-m_{2}^{2}$,
respectively. }
\label{fig:Fit}
\end{figure}

\begin{table}[tbp]
\begin{tabular}{|c|c|}
\hline\hline
Parameters & Values (in $\mathrm{MeV}$) \\ \hline\hline
$m_1[m_{J/\psi}]$ & $3096.900 \pm 0.006$ \\
$f_1[f_{J/\psi}]$ & $409 \pm 15$ \\
$m_{1}^{\ast}[m_{\psi^{\prime}}]$ & $3686.10 \pm 0.06$ \\
$f_{1}^{\ast}[f_{\psi^{\prime}}]$ & $279 \pm 8$ \\
$m_2[m_{\eta_c}]$ & $2983.9 \pm 0.4$ \\
$f_2[f_{\eta_c}]$ & $398.1 \pm 1.0$ \\
$m_{2}^{\ast}[m_{\eta_c(2S)}]$ & $3637.5 \pm 1.1$ \\
$f_{2}^{\ast}[f_{\eta_c(2S)}]$ & $331 $ \\
$m_3[m_{\chi _{c1}}]$ & $3510.67 \pm 0.05$ \\
$f_3[f_{\chi _{c1}}]$ & $344 \pm 27$ \\
$m_4[m_{\chi_{c0}}]$ & $3414.71 \pm 0.30$ \\
$f_4[f_{\chi_{c0}}]$ & $343 $ \\ \hline\hline
\end{tabular}%
\caption{Masses and decay constants of the various $\overline{c}c$ mesons
which have been used in numerical computations. }
\label{tab:Param}
\end{table}
The invariant amplitude $\Pi ^{\mathrm{OPE}}(p^{2},p^{\prime 2},q^{2})$
which corresponds to the term $\sim g_{\mu \nu }$ in Eq.\ (\ref{eq:QCDside})
forms the QCD side of the sum rule. Having equated the amplitudes $\Pi ^{%
\mathrm{OPE}}(p^{2},p^{\prime 2},q^{2})$ and $\Pi ^{\mathrm{Phys}%
}(p^{2},p^{\prime 2},q^{2})$ and performed the doubly Borel transforms and
continuum subtractions, one can find the sum rule for the form factor $%
g_{1}(q^{2})$%
\begin{equation}
g_{1}(q^{2})=\frac{2}{fmf_{1}^{2}m_{1}^{2}}\frac{q^{2}-m_{1}^{2}}{%
m^{2}-m_{1}^{2}-q^{2}}e^{m^{2}/M_{1}^{2}}e^{m_{1}^{2}/M_{2}^{2}}\Pi (\mathbf{%
M}^{2},\mathbf{s}_{0},q^{2}).  \label{eq:SRCoup}
\end{equation}%
Here,
\begin{equation}
\Pi (\mathbf{M}^{2},\mathbf{s}_{0},q^{2})=\int_{16m_{c}^{2}}^{s_{0}}ds%
\int_{4m_{c}^{2}}^{s_{0}^{\prime }}ds^{\prime }\rho (s,s^{\prime
},q^{2})e^{-s/M_{1}^{2}}e^{-s^{\prime }/M_{2}^{2}},  \label{eq:SCoupl}
\end{equation}%
is the function $\Pi ^{\mathrm{OPE}}(p^{2},p^{\prime 2},q^{2})$ after the
Borel transformations and subtraction procedures. It is expressed using the
spectral density $\rho (s,s^{\prime },q^{2})$: the latter is calculated as a
relevant imaginary part of $\Pi _{\mu \nu }^{\mathrm{OPE}}(p,p^{\prime })$.
In Eq.\ (\ref{eq:SCoupl}) $\mathbf{M}^{2}=(M_{1}^{2},M_{2}^{2})$ and $%
\mathbf{s}_{0}=(s_{0},s_{0}^{\prime })$ are the Borel and continuum
threshold parameters, respectively.

The form factor $g_{1}(q^{2})$ depends on the masses and decay constants of
the molecule $\mathcal{M}_{1}$ and meson $J/\psi $ , which are input
parameters in calculations. Their numerical values are collected in Table\ %
\ref{tab:Param}. Additionally, this table contains the parameters of the $%
\psi ^{\prime }$, $\eta _{c}$, $\eta _{c}(2S)$, $\chi _{c1}(1P)$, and $\chi
_{c0}(1P)$ mesons which are necessary to explore decay modes of the hadronic
molecules $\mathcal{M}_{1}$ and $\mathcal{M}_{2}$. For the masses of the
particles, we utilize information from Ref.\ \cite{PDG:2022}. For decay
constant of the meson $J/\psi $, we employ the experimental value from Ref.\
\cite{Kiselev:2001xa}. As $f_{\eta _{c}}$ we use results of the QCD lattice
simulations \cite{Hatton:2020qhk}, whereas for $f_{\chi _{c1}}$ and $f_{\chi
_{c0}}$ -- the sum rule predictions from Refs.\ \cite%
{VeliVeliev:2012cc,Veliev:2010gb}.

To carry out numerical computations, it is also necessary to choose the
working regions for the parameters $\mathbf{M}^{2}$ and $\mathbf{s}_{0}$.
For $M_{1}^{2}$ and $s_{0}$, associated with the $\mathcal{M}_{1}$ channel,
we apply the working windows of Eq.\ (\ref{eq:Wind1}). The parameters $%
(M_{2}^{2},\ s_{0}^{\prime })$ for the $J/\psi $ channel are varied inside
the intervals%
\begin{equation}
M_{2}^{2}\in \lbrack 4,5]~\mathrm{GeV}^{2},\ s_{0}^{\prime }\in \lbrack
12,13]~\mathrm{GeV}^{2}.  \label{eq:Wind3}
\end{equation}

It is a fact that the sum rule approach leads to reliable predictions in the
deep-Euclidean region $q^{2}<0$. For our purposes, it is suitable to
introduce a new variable $Q^{2}=-q^{2}$ and present the obtained function by
$g_{1}(Q^{2})$. The interval of $Q^{2}$ studied by the sum rule analysis
contains the region $Q^{2}=1-10~\mathrm{GeV}^{2}$. The results of analyses
are plotted in Fig.\ \ref{fig:Fit}.

But the width of the decay $\mathcal{M}_{1}\rightarrow J/\psi J/\psi $ is
determined by the the form factor $g_{1}(q^{2})$ at the mass shell $%
q^{2}=m_{1}^{2}$, i.e., one has to find $g_{1}(Q^{2}=-m_{1}^{2})$. To
overcome this problem, we use a fit function $\mathcal{G}_{1}(Q^{2})$, which
at momenta $Q^{2}>0$ gives the same values as the sum rule predictions, but
it can be extrapolated to the region of $Q^{2}<0$. In this paper, we employ
the functions $\mathcal{G}_{i}(Q^{2})$
\begin{equation}
\mathcal{G}_{i}(Q^{2})=\mathcal{G}_{i}^{0}\mathrm{\exp }\left[ c_{i}^{1}%
\frac{Q^{2}}{m^{2}}+c_{i}^{2}\left( \frac{Q^{2}}{m^{2}}\right) ^{2}\right]
\label{eq:FitF}
\end{equation}%
with parameters $\mathcal{G}_{i}^{0}$, $c_{i}^{1}$, and $c_{i}^{2}$.

Calculations demonstrate that $\mathcal{G}_{1}^{0}=3.41~\mathrm{GeV}^{-1}$, $%
c_{1}^{1}=2.18$, and $c_{1}^{2}=-2.21$ lead to reasonable agreement with the
sum rule's data for $g_{1}(Q^{2})$ depicted in Fig.\ \ref{fig:Fit}. At the
mass shell $q^{2}=m_{1}^{2}$ the function $\mathcal{G}_{1}(Q^{2})$ equals to
\begin{equation}
g_{1}\equiv \mathcal{G}_{1}(-m_{1}^{2})=(1.75\pm 0.41)\ \mathrm{GeV}^{-1}.
\end{equation}%
The partial width of the process $\mathcal{M}_{1}\rightarrow J/\psi J/\psi $
can be obtained by means of the following formula
\begin{equation}
\Gamma \left[ \mathcal{M}_{1}\rightarrow J/\psi J/\psi \right] =g_{1}^{2}%
\frac{\lambda _{1}}{8\pi }\left( \frac{m_{1}^{4}}{m^{2}}+\frac{2\lambda
_{1}^{2}}{3}\right) ,  \label{eq:PartDW}
\end{equation}%
where $\lambda _{1}=\lambda (m,m_{1},m_{1})$ and
\begin{equation}
\lambda (a,b,c)=\frac{\sqrt{%
a^{4}+b^{4}+c^{4}-2(a^{2}b^{2}+a^{2}c^{2}+b^{2}c^{2})}}{2a}.
\end{equation}

It is easy to find that
\begin{equation}
\Gamma \left[ \mathcal{M}_{1}\rightarrow J/\psi J/\psi \right] =(142\pm 47)~%
\mathrm{MeV}.  \label{eq:DW1}
\end{equation}


\subsubsection{Process $\mathcal{M}_{1}\rightarrow \protect\eta _{c}\protect%
\eta _{c}$}


The process $\mathcal{M}_{1}\rightarrow \eta _{c}\eta _{c}$ is another decay
channel of the hadronic molecule $\mathcal{M}_{1}$. Investigation of this
process runs, with some modifications, in accordance with the scheme
explained above. The strong coupling $g_{2}$ that describes the vertex $%
\mathcal{M}_{1}\eta _{c}\eta _{c}$ is extracted from the correlation
function
\begin{equation}
\Pi (p,p^{\prime })=i^{2}\int d^{4}xd^{4}ye^{ip^{\prime }y}e^{-ipx}\langle 0|%
\mathcal{T}\{J^{\eta _{c}}(y)J^{\eta _{c}}(0)J^{\dagger }(x)\}|0\rangle ,
\label{eq:CF4}
\end{equation}%
where
\begin{equation}
J^{\eta _{c}}(x)=\overline{c}_{i}(x)i\gamma _{5}c_{i}(x),  \label{eq:CR3}
\end{equation}%
is the interpolating current for the meson $\eta _{c}$.

The physical side of the sum rule for the form factor $g_{2}(q^{2})$ is
derived by separating the contribution of the ground-state and the effects
of the higher states and continuum from each other. Then, the correlation
function (\ref{eq:CF4}) can be presented in the following form%
\begin{eqnarray}
&&\Pi ^{\mathrm{Phys}}(p,p^{\prime })=\frac{\langle 0|J^{\eta _{c}}|\eta
_{c}(p^{\prime })\rangle }{p^{\prime 2}-m_{2}^{2}}\frac{\langle 0|J^{\eta
_{c}}|\eta _{c}(q)\rangle }{q^{2}-m_{2}^{2}}\langle \eta _{c}(p^{\prime
})\eta _{c}(q)|\mathcal{M}_{1}(p)\rangle  \notag \\
&&\times \frac{\langle \mathcal{M}_{1}(p)|J^{\dagger }|0\rangle }{p^{2}-m^{2}%
}+\cdots ,  \label{eq:CF5}
\end{eqnarray}%
with $m_{2}$ being the mass of the $\eta _{c}$ meson.

We define the vertex composed of a scalar and two pseudoscalar particles by
means of the formula
\begin{equation}
\langle \eta _{c}(p^{\prime })\eta _{c}(q)|\mathcal{M}_{1}(p)\rangle
=g_{2}(q^{2})p\cdot p^{\prime }.  \label{eq:ME5}
\end{equation}%
To rewrite the correlator $\Pi ^{\mathrm{Phys}}(p,p^{\prime })$ in terms of
physical parameters of particles $\mathcal{M}_{1}$ and $\eta _{c}$, we also
use the matrix elements Eq.\ (\ref{eq:ME1}) and
\begin{equation}
\langle 0|J^{\eta _{c}}|\eta _{c}\rangle =\frac{f_{2}m_{2}^{2}}{2m_{c}},
\label{eq:ME4}
\end{equation}%
where $f_{2}$ is the decay constant of the $\eta _{c}$ meson. The
correlation function $\Pi ^{\mathrm{Phys}}(p,p^{\prime })$ then becomes
equal to
\begin{equation}
\Pi ^{\mathrm{Phys}}(p,p^{\prime })=g_{2}(q^{2})\frac{fmf_{2}^{2}m_{2}^{4}}{%
4m_{c}^{2}\left( p^{2}-m^{2}\right) \left( p^{\prime 2}-m_{2}^{2}\right) }%
\frac{m^{2}+m_{2}^{2}-q^{2}}{2(q^{2}-m_{2}^{2})}+\cdots .  \label{eq:CF6}
\end{equation}%
The function $\Pi ^{\mathrm{Phys}}(p,p^{\prime })$ has a Lorentz structure
which is proportional to $\mathrm{I}$, hence right-hand side of Eq.\ (\ref%
{eq:CF6}) is the corresponding invariant amplitude $\widehat{\Pi }^{\mathrm{%
Phys}}(p^{2},p^{\prime 2},q^{2})$.

We also find the function $\Pi ^{\mathrm{OPE}}(p,p^{\prime })$
\begin{eqnarray}
&&\Pi ^{\mathrm{OPE}}(p,p^{\prime })=2i^{2}\int d^{4}xd^{4}ye^{ip^{\prime
}y}e^{-ipx}\left\{ \mathrm{Tr}\left[ \gamma _{5}S_{c}^{ia}(y-x)\gamma
_{5}S_{c}^{ai}(x-y)\right] \right.  \notag \\
&& \left. \times \mathrm{Tr}\left[ \gamma _{5}S_{c}^{jb}(-x)\gamma
_{5}S_{c}^{bj}(x){}\right] -\mathrm{Tr}\left[ \gamma
_{5}S_{c}^{ia}(y-x)\gamma _{5}S_{c}^{aj}(x){}\gamma _{5}S_{c}^{jb}(-x)\gamma
_{5}S_{c}^{bi}(x-y)\right] \right\},  \label{eq:QCDside2}
\end{eqnarray}%
and denote the amplitude corresponding to a structure $\sim \mathrm{I}$ by $%
\widehat{\Pi }^{\mathrm{OPE}}(p^{2},p^{\prime 2},q^{2})$. The sum rule for
the strong form factor $g_{2}(q^{2})$ reads%
\begin{equation}
g_{2}(q^{2})=\frac{8m_{c}^{2}}{fmf_{2}^{2}m_{2}^{4}}\frac{q^{2}-m_{2}^{2}}{%
m^{2}+m_{2}^{2}-q^{2}}e^{m^{2}/M_{1}^{2}}e^{m_{2}^{2}/M_{2}^{2}}\widehat{\Pi
}(\mathbf{M}^{2},\mathbf{s}_{0},q^{2}).  \label{eq:SRCoup2}
\end{equation}

Numerical computations are carried out using the parameters of the meson $%
\eta _{c}$ from Table\ \ref{tab:Param}. The Borel and continuum subtraction
parameters $M_{1}^{2}$ and $s_{0}$ in the $\mathcal{M}_{1}$ channel is
chosen as in Eq.\ (\ref{eq:Wind1}), whereas for $M_{2}^{2}$ and $%
s_{0}^{\prime }$ that correspond to the $\eta _{c}$ channel, we employ
\begin{equation}
M_{2}^{2}\in \lbrack 3.5,4.5]~\mathrm{GeV}^{2},\ s_{0}^{\prime }\in \lbrack
11,12]~\mathrm{GeV}^{2}.  \label{eq:Wind4}
\end{equation}

The fit function $\mathcal{G}_{2}(Q^{2})$ has the following parameters: $%
\mathcal{G}_{2}^{0}=1.52~\mathrm{GeV}^{-1}$, $c_{2}^{1}=2.84$, and $%
c_{2}^{2}=-2.73$. For the strong coupling $g_{2}$, we get
\begin{equation}
g_{2}\equiv \mathcal{G}_{2}(-m_{2}^{2})=(6.9\pm 1.5)\times 10^{-1}\ \mathrm{%
GeV}^{-1}.
\end{equation}%
The width of the process $\mathcal{M}_{1}\rightarrow \eta _{c}\eta _{c}$ is
determined by means of the formula%
\begin{equation}
\Gamma \left[ \mathcal{M}_{1}\rightarrow \eta _{c}\eta _{c}\right] =g_{2}^{2}%
\frac{m_{2}^{2}\lambda _{2}}{8\pi }\left( 1+\frac{\lambda _{2}^{2}}{m_{2}^{2}%
}\right),  \label{eq:PDw2}
\end{equation}%
where $\lambda _{2}=\lambda (m,m_{2},m_{2})$. Finally, we obtain
\begin{equation}
\Gamma \left[ \mathcal{M}_{1}\rightarrow \eta _{c}\eta _{c}\right] =(178\pm
55)~\mathrm{MeV}.  \label{eq:DW2}
\end{equation}%
The parameters of the decays $\mathcal{M}_{1}\rightarrow J/\psi J/\psi $ and
$\mathcal{M}_{1}\rightarrow \eta _{c}\eta _{c}$ are shown in Table \ref%
{tab:Channels}.

Based on these results, it is not difficult to find that
\begin{equation}
\Gamma _{\mathcal{M}_{1}}=(320\pm 72)~\mathrm{MeV},  \label{eq:FW}
\end{equation}%
which is the full width of the hadronic molecule $\mathcal{M}_{1}$.

\begin{table}[tbp]
\begin{tabular}{|c|c|c|c|}
\hline\hline
$i$ & Channels & $g_{i}^{(\ast)}\times 10~(\mathrm{GeV}^{-1})$ & $%
\Gamma_{i}~(\mathrm{MeV})$ \\ \hline
$1$ & $\mathcal{M}_{1}\to J/\psi J/\psi$ & $17.5 \pm 4.1$ & $142 \pm 47$ \\
$2$ & $\mathcal{M}_{1}\to \eta_{c}\eta_{c}$ & $6.9 \pm 1.5$ & $178 \pm 55$
\\ \hline
$3$ & $\mathcal{M}_{2}\to J/\psi J\psi$ & $4.1 \pm 0.8$ & $38 \pm 11$ \\
$3^{\ast}$ & $\mathcal{M}_{2}\to J/\psi\psi^{\prime}$ & $3.3 \pm 0.7$ & $11
\pm 4$ \\
$4$ & $\mathcal{M}_{2}\to \eta_{c}\eta_{c}$ & $2.1 \pm 0.4$ & $39 \pm 11$ \\
$4^{\ast}$ & $\mathcal{M}_{2}\to \eta_{c}\eta_{c}(2S)$ & $1.34 \pm 0.26$ & $%
12 \pm 4$ \\
$5$ & $\mathcal{M}_{2}\to \eta_{c}\chi_{c1}$ & $29 \pm 6^{\star} $ & $16 \pm
5$ \\
$6$ & $\mathcal{M}_{2} \to \chi_{c0}\chi_{c0}$ & $2.7 \pm 0.43 $ & $22 \pm 5
$ \\ \hline\hline
\end{tabular}%
\caption{Decay channels of the molecules $\mathcal{M}_{1(2)}$, strong
couplings $g_{i}^{(\ast)}$, and partial widths $\Gamma _{i}$. The
star-labeled coupling $g_{5}$ is dimensionless.}
\label{tab:Channels}
\end{table}


\section{Hadronic molecule $\protect\chi _{c0}\protect\chi _{c0}$}

\label{sec:MW2}

This part of the article is devoted to thorough investigations of the
molecule $\mathcal{M}_{2}=\chi _{c0}\chi _{c0}$ which imply calculations of
the mass $\widetilde{m}$ and coupling $\widetilde{f}$, as well as the full
width $\Gamma _{\mathcal{M}_{2}}$ of this compound by employing its numerous
decay modes.\newline


\subsection{Spectroscopic parameters $\widetilde{m}$ and $\widetilde{f}$}


In the case of the molecule $\mathcal{M}_{2}=\chi _{c0}\chi _{c0}$ the
two-point correlation function that should be analyzed has the form
\begin{equation}
\widetilde{\Pi }(p)=i\int d^{4}xe^{ipx}\langle 0|\mathcal{T}\{\widetilde{J}%
(x)\widetilde{J}^{\dag }(0)\}|0\rangle,  \label{eq:CF1A}
\end{equation}%
where $\widetilde{J}(x)$ is the interpolating current for the molecule $%
\mathcal{M}_{2}$. We treat $\mathcal{M}_{2}$ as a hadronic state built of
the scalar mesons $\chi _{c0}\chi _{c0}$, therefore define the relevant
interpolating current as
\begin{equation}
\widetilde{J}(x)=\overline{c}_{a}(x)c_{a}(x)\overline{c}_{b}(x)c_{b}(x).
\label{eq:CR1A}
\end{equation}%
The physical side of the sum rule
\begin{equation}
\widetilde{\Pi }^{\mathrm{Phys}}(p)=\frac{\widetilde{f}^{2}\widetilde{m}^{2}%
}{\widetilde{m}^{2}-p^{2}}+\cdots,  \label{eq:Phen3}
\end{equation}%
does not differ from Eq.\ (\ref{eq:Phen2}), but $\widetilde{m}$ and $%
\widetilde{f}$ are now the mass and coupling of the molecule $\mathcal{M}%
_{2} $ introduced through the matrix element
\begin{equation}
\langle 0|\widetilde{J}|\mathcal{M}_{2}(p)\rangle =\widetilde{f}\widetilde{m}%
.  \label{eq:ME1A}
\end{equation}%
The amplitude $\widetilde{\Pi }^{\mathrm{Phys}}(p^{2})$ required for the
following analysis is given by the expression in the right-hand side of Eq.\
(\ref{eq:Phen3}).

The correlation function $\widetilde{\Pi }^{\mathrm{OPE}}(p)$ computed using
the $c$-quark propagators is written down in Eq.\ (\ref{eq:QCD2}). The sum
rules for $\widetilde{m}$ and $\widetilde{f}$ are determined by Eqs.\ (\ref%
{eq:Mass}) and (\ref{eq:Coupl}) with evident replacements.

The working windows for the Borel and continuum subtraction parameters $%
M^{2} $ and $s_{0}$ are:
\begin{equation}
M^{2}\in \lbrack 5.5,7]~\mathrm{GeV}^{2},\ s_{0}\in \lbrack 54,55]~\mathrm{%
GeV}^{2}.  \label{eq:Wind2}
\end{equation}%
At $M^{2}=5.5~\mathrm{GeV}^{2}$ and $7~\mathrm{GeV}^{2}$ the pole
contribution constitutes $\mathrm{PC}=0.75$ and $0.48$ parts of the
correlation function, respectively. The pole contribution changes within
limits
\begin{equation}
0.75\geq \mathrm{PC}\geq 0.48.
\end{equation}%
On average in $s_{0}$ the pole contribution is $\mathrm{PC}\geq 0.5$. The
dimension-$4$ term is negative and at $M^{2}=5.5~\mathrm{GeV}^{2}$
constitutes $\simeq 19.9\%$ of the correlator.

\begin{figure}[h!]
\begin{center}
\includegraphics[totalheight=6cm,width=8cm]{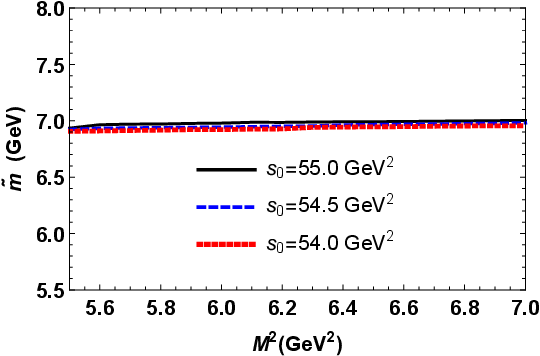}%
\includegraphics[totalheight=6cm,width=8cm]{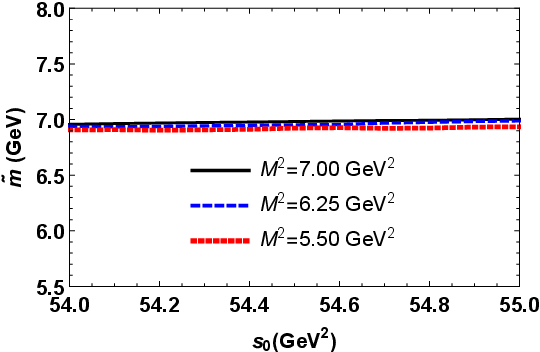}
\end{center}
\caption{ Dependence of the $\mathcal{M}_2$ molecule's mass $\widetilde{m}$
on the parameters $M^2$ (left), and $s_0$ (right).}
\label{fig:MassB}
\end{figure}

The mass and current coupling of $\mathcal{M}_{2}$ are:
\begin{equation}
\widetilde{m}=(6954\pm 50)~\mathrm{MeV},\ \ \widetilde{f}=(1.71\pm
0.12)\times 10^{-2}~\mathrm{GeV}^{4}.  \label{eq:Result2}
\end{equation}%
These results are obtained as mean values of $\widetilde{m}$ and $\widetilde{%
f}$ averaged over the regions Eq.\ (\ref{eq:Wind2}). They effectively
correspond to the sum rule predictions at the point $M^{2}=6.2~\mathrm{GeV}%
^{2}$ and $s_{0}=54.5~\mathrm{GeV}^{2}$, where $\mathrm{PC}=0.64$. In Fig.\ %
\ref{fig:MassB}, we plot $\widetilde{m}$ as a function of $M^{2}$ and $s_{0}$%
.

\subsection{Decays of $\mathcal{M}_{2}$}


The prediction for the mass of the molecule $\mathcal{M}_{2}$ determines its
kinematically allowed decay channels. First of all, they are processes $%
\mathcal{M}_{2}\rightarrow J/\psi J/\psi $ and $\mathcal{M}_{2}\rightarrow
J/\psi \psi ^{\prime }$. The $\widetilde{m}$ satisfies kinematical
restrictions for productions of $\eta _{c}\eta _{c}$ and $\eta _{c}\eta
_{c}(2S)$ pairs. The $\mathcal{M}_{2}$ can decay to $\eta _{c}\chi _{c1}(1P)$
and $\chi _{c0}\chi _{c0}$ mesons as well. The decay $\mathcal{M}%
_{2}\rightarrow \eta _{c}\chi _{c1}(1P)$ is the $P$-wave process, whereas
other ones are $S$-wave modes.


\subsubsection{$\mathcal{M}_{2}\rightarrow J/\protect\psi J/\protect\psi $
and $\mathcal{M}_{2}\rightarrow J/\protect\psi \protect\psi ^{\prime }$}


The three-point sum rules for the strong form factors $g_{3}(q^{2})$ and $%
g_{3}^{\ast }(q^{2})$ which describe interaction of particles at vertices $%
\mathcal{M}_{2}J/\psi J/\psi $ \ and $\mathcal{M}_{2}J/\psi \psi ^{\prime }$
respectively, can be extracted from studies of the correlation function
\begin{equation}
\widetilde{\Pi }_{\mu \nu }(p,p^{\prime })=i^{2}\int
d^{4}xd^{4}ye^{ip^{\prime }y}e^{-ipx}\langle 0|\mathcal{T}\{J_{\mu }^{\psi
}(y)J_{\nu }^{\psi }(0)\widetilde{J}^{\dagger }(x)\}|0\rangle .
\label{eq:CF2A}
\end{equation}

Firstly, we express $\widetilde{\Pi }_{\mu \nu }(p,p^{\prime })$ using the
physical parameters of particles involved in these decays. The molecule $%
\mathcal{M}_{2}$ can decay to $J/\psi J/\psi $ and $J/\psi \psi ^{\prime }$
mesons, therefore in $\widetilde{\Pi }_{\mu \nu }(p,p^{\prime })$ we isolate
contributions of the particles $J/\psi $ and $\psi ^{\prime }$ from effects
of higher resonances and continuum states. Then, the physical side of the
sum rule $\widetilde{\Pi }_{\mu \nu }^{\mathrm{Phys}}(p,p^{\prime })$ is
determined by Eq.\ (\ref{eq:CF3A}). It can be rewritten using the matrix
elements of the particles $\mathcal{M}_{2}$, $J/\psi $ and $\psi ^{\prime }$
\begin{eqnarray}
&&\widetilde{\Pi }_{\mu \nu }^{\mathrm{Phys}}(p,p^{\prime })=g_{3}(q^{2})%
\frac{\widetilde{f}\widetilde{m}f_{1}^{2}m_{1}^{2}}{\left( p^{2}-\widetilde{m%
}^{2}\right) \left( p^{\prime 2}-m_{1}^{2}\right) (q^{2}-m_{1}^{2})}\left[
\frac{1}{2}\left( \widetilde{m}^{2}-m_{1}^{2}-q^{2}\right) g_{\mu \nu
}-q_{\mu }p_{\nu }^{\prime }\right]  \notag \\
&&+g_{3}^{\ast }(q^{2})\frac{\widetilde{f}\widetilde{m}f_{1}m_{1}f_{1}^{\ast
}m_{1}^{\ast }}{\left( p^{2}-\widetilde{m}^{2}\right) \left( p^{\prime
2}-m_{1}^{\ast 2}\right) (q^{2}-m_{1}^{2})}\left[ \frac{1}{2}\left(
\widetilde{m}^{2}-m_{1}^{\ast 2}-q^{2}\right) g_{\mu \nu }-q_{\mu }p_{\nu
}^{\prime }\right] +\cdots ,  \notag \\
&&  \label{eq:CorrF5}
\end{eqnarray}%
where $m_{1}^{\ast }$ and $f_{1}^{\ast }$ are the mass and decay constant of
the meson $\psi ^{\prime }$. In what follows, we use the component of $%
\widetilde{\Pi }_{\mu \nu }^{\mathrm{Phys}}(p,p^{\prime })$ that is
proportional to $g_{\mu \nu }$, and denote the relevant invariant amplitude
by $\widetilde{\Pi }^{\mathrm{Phys}}(p^{2},p^{\prime 2},q^{2})$.

The second term of the sum rules, \ i.e., the correlation function $%
\widetilde{\Pi }_{\mu \nu }^{\mathrm{OPE}}(p,p^{\prime })$ is presented in
Eq.\ (\ref{eq:QCD3}). An amplitude $\widetilde{\Pi }^{\mathrm{OPE}%
}(p^{2},p^{\prime 2},q^{2})$ which corresponds to the term $\sim g_{\mu \nu
} $ in $\widetilde{\Pi }_{\mu \nu }^{\mathrm{OPE}}(p,p^{\prime })$
establishes the QCD side of the sum rules. By equating the amplitudes $%
\widetilde{\Pi }^{\mathrm{OPE}}(p^{2},p^{\prime 2},q^{2})$ and $\widetilde{%
\Pi }^{\mathrm{Phys}}(p^{2},p^{\prime 2},q^{2})$, applying the the Borel
transformations and carrying out continuum subtractions, one can find the
sum rules for the form factors $g_{3}(q^{2})$ and $g_{3}^{\ast }(q^{2})$.
Let us note after these manipulations the $\widetilde{\Pi }^{\mathrm{OPE}%
}(p^{2},p^{\prime 2},q^{2})$ takes the form Eq.\ (\ref{eq:SCoupl}) with a
new spectral density $\widetilde{\rho }(s,s^{\prime },q^{2})$.

To find the form factors $g_{3}(q^{2})$ and $g_{3}^{\ast }(q^{2})$, we keep
the following approach. At the first phase, we compute the form factor $%
g_{3}(q^{2})$ by choosing in the $J/\psi $ channel $4m_{c}^{2}<s_{0}^{\prime
}<m_{1}^{\ast 2}$. This means that, we exclude the second term in Eq.\ (\ref%
{eq:CorrF5}) from analysis by including it into higher resonances and
continuum states. As a result, the physical side of the sum rule contains a
contribution coming only from ground-level states. The sum rule for the form
factor $g_{3}(q^{2})$ is determined by Eq.\ (\ref{eq:SRCoup}) after
substitutions $\Pi (\mathbf{M}^{2},\mathbf{s}_{0},q^{2})\rightarrow
\widetilde{\Pi }(\mathbf{M}^{2},\mathbf{s}_{0},q^{2})$ and $fm\rightarrow
\widetilde{f}\widetilde{m}$. At the second stage of computations, we fix $%
s_{0}^{\ast \prime }>$ $m_{1}^{\ast 2}$, and take into account the second
term in Eq.\ (\ref{eq:CorrF5}). Afterwards, using results obtained for $%
g_{3}(q^{2})$, we determine $g_{3}^{\ast }(q^{2})$.

In numerical computations the working regions for $M_{1}^{2}$ and $s_{0}$ in
the $\mathcal{M}_{2}$ channel are chosen as in Eq.\ (\ref{eq:Wind2}). The
parameters $(M_{2}^{2},\ s_{0}^{\prime })$ for the $J/\psi $ channel are
changed within limits given by Eq.\ (\ref{eq:Wind3}). The sum rule
calculations are carried out in deep-Euclidean region $q^{2}=-(1\div 10)~%
\mathrm{GeV}^{2}$. The fit function $\mathcal{G}_{3}(Q^{2})$ necessary to
extrapolate these data to region of $q^{2}>0$ has the parameters $\mathcal{G}%
_{3}^{0}=0.87~\mathrm{GeV}^{-1}$, $c_{3}^{1}=3.03$, and $c_{3}^{2}=-3.64$.
At the mass shell $q^{2}=m_{1}^{2}$ this function determines the strong
coupling $g_{3}$
\begin{equation}
g_{3}\equiv \mathcal{G}_{3}(-m_{1}^{2})=(4.1\pm 0.8)\times 10^{-1}\ \mathrm{%
GeV}^{-1}.  \label{eq:Coup1}
\end{equation}%
Partial width of the process $\mathcal{M}_{2}\rightarrow J/\psi J/\psi $ can
be found by means of Eq.\ (\ref{eq:PartDW}) after substitutions $%
g_{1}\rightarrow g_{3}$, $m^{2}\rightarrow \widetilde{m}^{2}$ and $\lambda
_{1}\rightarrow \lambda _{3}=\lambda (\widetilde{m},m_{1},m_{1})$. It is not
difficult to get
\begin{equation}
\Gamma \left[ \mathcal{M}_{2}\rightarrow J/\psi J/\psi \right] =(38\pm 11)~%
\mathrm{MeV}.  \label{eq:DW1AA}
\end{equation}

The process $\mathcal{M}_{2}\rightarrow J/\psi \psi ^{\prime }$ can be
studied in accordance with a scheme described above. In this phase of the
analysis, in the $\psi ^{\prime }$ channel, we employ
\begin{equation}
M_{2}^{2}\in \lbrack 4,5]~\mathrm{GeV}^{2},\ s_{0}^{\ast \prime }\in \lbrack
15,16]~\mathrm{GeV}^{2}.
\end{equation}%
It is worth noting that $s_{0}^{\ast \prime }$ is limited by the mass $%
m(3S)=4039~\mathrm{MeV}$ of the charmonium $\psi (3S)$ \cite{PDG:2022}. For
this decay, the extrapolating function $\mathcal{G}_{3}^{\ast }(Q^{2})$ has
the parameters $\mathcal{G}_{3}^{0\ast }=0.68~\mathrm{GeV}^{-1}$, $%
c_{3}^{1\ast }=2.90$, and $c_{3}^{2\ast }=-3.54$. The strong coupling $%
g_{3}^{\ast }$ is calculated at the mass shell $q^{2}=m_{1}^{2}$
\begin{equation}
g_{3}^{\ast }\equiv \mathcal{G}_{3}^{\ast }(-m_{1}^{2})=(3.3\pm 0.7)\times
10^{-1}\ \mathrm{GeV}^{-1}.  \label{eq:Coup2}
\end{equation}%
The partial width of the decay $\mathcal{M}_{2}\rightarrow J/\psi \psi
^{\prime }$ is
\begin{equation}
\Gamma \left[ \mathcal{M}_{2}\rightarrow J/\psi \psi ^{\prime }\right]
=(11\pm 4)~\mathrm{MeV}.  \label{eq:DW1A}
\end{equation}


\subsubsection{$\mathcal{M}_{2}\rightarrow \protect\eta _{c}\protect\eta %
_{c} $ and $\mathcal{M}_{2}\rightarrow \protect\eta _{c}\protect\eta %
_{c}(2S) $}


The processes $\mathcal{M}_{2}\rightarrow \eta _{c}\eta _{c}$ and $\mathcal{M%
}_{2}\rightarrow \eta _{c}\eta _{c}(2S)$ can be investigated by the similar
manner. The strong couplings $g_{4}$ and $g_{4}^{\ast }$ that correspond to
the vertices $\mathcal{M}_{2}\eta _{c}\eta _{c}$ and $\mathcal{M}_{2}\eta
_{c}\eta _{c}(2S)$ can be extracted from the correlation function
\begin{equation}
\widetilde{\Pi }(p,p^{\prime })=i^{2}\int d^{4}xd^{4}ye^{ip^{\prime
}y}e^{-ipx}\langle 0|\mathcal{T}\{J^{\eta _{c}}(y)J^{\eta _{c}}(0)\widetilde{%
J}^{\dagger }(x)\}|0\rangle .  \label{eq:CF4A}
\end{equation}%
Separating the ground-level and first excited state contributions from
effects of higher resonances and continuum states, we can write the
correlation function $\widetilde{\Pi }^{\mathrm{Phys}}(p,p^{\prime })$ which
is determined by Eq.\ (\ref{eq:CF5A}). It can be further simplified using
known matrix elements and takes the form%
\begin{eqnarray}
&&\widetilde{\Pi }^{\mathrm{Phys}}(p,p^{\prime })=g_{4}(q^{2})\frac{%
\widetilde{f}\widetilde{m}f_{2}^{2}m_{2}^{4}}{8m_{c}^{2}\left( p^{2}-%
\widetilde{m}^{2}\right) \left( p^{\prime 2}-m_{2}^{2}\right) }\frac{%
\widetilde{m}^{2}+m_{2}^{2}-q^{2}}{q^{2}-m_{2}^{2}}  \notag \\
&&+g_{4}^{\ast }(q^{2})\frac{\widetilde{f}\widetilde{m}f_{2}m_{2}^{2}f_{2}^{%
\ast }m_{2}^{\ast 2}}{8m_{c}^{2}\left( p^{2}-\widetilde{m}^{2}\right) \left(
p^{\prime 2}-m_{2}^{\ast 2}\right) }\frac{\widetilde{m}^{2}+m_{2}^{\ast
2}-q^{2}}{q^{2}-m_{2}^{2}}+\cdots .  \label{eq:CF6A}
\end{eqnarray}%
where $m_{2}^{\ast }$ and $f_{2}^{\ast }$ are the mass and decay constant of
the $\eta _{c}(2S)$ meson. The correlation function $\widetilde{\Pi }^{%
\mathrm{Phys}}(p,p^{\prime })$ has simple Lorentz structure proportional to $%
\mathrm{I}$, hence right-hand side of Eq.\ (\ref{eq:CF6A}) is the
corresponding invariant amplitude $\widetilde{\Pi }^{\mathrm{Phys}%
}(p^{2},p^{\prime 2},q^{2})$.

The QCD side of the sum rule $\widetilde{\Pi }^{\mathrm{OPE}}(p,p^{\prime })$
is given by Eq.\ (\ref{eq:QCD4}). The sum rule for the strong form factor $%
g_{4}(q^{2})$ is determined by Eq.\ (\ref{eq:SRCoup2}) with replacements $%
fm\rightarrow \widetilde{f}\widetilde{m}$ and $\widehat{\Pi }\rightarrow
\widetilde{\Pi }$, where $\widetilde{\Pi }(\mathbf{M}^{2},\mathbf{s}%
_{0},q^{2})$ corresponds to the correlation function $\widetilde{\Pi }^{%
\mathrm{OPE}}(p,p^{\prime })$.

Numerical computations are carried out using Eq.\ (\ref{eq:SRCoup2}),
parameters of the meson $\eta _{c}$ from Table\ \ref{tab:Param}, and working
regions for $\mathbf{M}^{2}$ and $\mathbf{s}_{0}$. The Borel and continuum
subtraction parameters $M_{1}^{2}$ and $s_{0}$ in the $\mathcal{M}_{2}$
channel are chosen as in Eq.\ (\ref{eq:Wind2}), whereas for $M_{2}^{2}$ and $%
s_{0}^{\prime }$ which correspond to the $\eta _{c}$ channel, we employ Eq.\
(\ref{eq:Wind4}).

The interpolating function $\mathcal{G}_{4}(Q^{2})$ necessary to determine
the coupling $g_{4}$ has the parameters: $\mathcal{G}_{4}^{0}=0.48~\mathrm{%
GeV}^{-1}$, $c_{4}^{1}=3.65$, and $c_{4}^{2}=-4.24$. For the strong coupling
$g_{4}$, we get
\begin{equation}
g_{4}\equiv \mathcal{G}_{4}(-m_{2}^{2})=(2.1\pm 0.4)\times 10^{-1}\ \mathrm{%
GeV}^{-1}.
\end{equation}%
The width of the process $\mathcal{M}_{2}\rightarrow \eta _{c}\eta _{c}$ is
determined by means of the formula Eq.\ (\ref{eq:PDw2}) with substitutions $%
g_{2}\rightarrow g_{4}$, $\lambda _{2}\rightarrow \lambda _{4}=\lambda (%
\widetilde{m},m_{2},m_{2})$. Our computations yield
\begin{equation}
\Gamma \left[ \mathcal{M}_{2}\rightarrow \eta _{c}\eta _{c}\right] =(39\pm
11)~\mathrm{MeV}.  \label{eq:DW2A}
\end{equation}%
For the channel $\mathcal{M}_{2}\rightarrow \eta _{c}\eta _{c}(2S)$, we use
\begin{equation}
M_{2}^{2}\in \lbrack 3.5,4.5]~\mathrm{GeV}^{2},\ s_{0}^{\ast \prime }\in
\lbrack 13,14]~\mathrm{GeV}^{2},  \label{eq:Wind4A}
\end{equation}%
and find
\begin{equation}
g_{4}^{\ast }\equiv \mathcal{G}_{4}^{\ast }(-m_{2}^{2})=(1.34\pm 0.26)\times
10^{-1}\ \mathrm{GeV}^{-1}.
\end{equation}%
The $g_{4}^{\ast }$ is evaluated using the fit function $\mathcal{G}%
_{4}^{\ast }(Q^{2})$ with the parameters $\mathcal{G}_{4}^{0\ast }=0.32~%
\mathrm{GeV}^{-1}$, $c_{4}^{1\ast }=3.64$, and $c_{4}^{2\ast }=-4.23$. The
width of this decay is equal to
\begin{equation}
\Gamma \left[ \mathcal{M}_{2}\rightarrow \eta _{c}\eta _{c}(2S)\right]
=(12\pm 4)~\mathrm{MeV}.
\end{equation}


\subsubsection{$\mathcal{M}_{2}\rightarrow \protect\eta _{c}\protect\chi %
_{c1}(1P)$ and $\mathcal{M}_{2}\rightarrow \protect\chi _{c0}\protect\chi %
_{c0}$}


Analysis of the $P$-wave process $\mathcal{M}_{2}\rightarrow \eta _{c}\chi
_{c1}(1P)$ is performed by the standard method. The three-point correlator
that should be studied in this case is
\begin{equation}
\widetilde{\Pi }_{\mu }(p,p^{\prime })=i^{2}\int d^{4}xd^{4}ye^{ip^{\prime
}y}e^{-ipx}\langle 0|\mathcal{T}\{J_{\mu }^{\chi _{c1}}(y)J^{\eta _{c}}(0)%
\widetilde{J}^{\dagger }(x)\}|0\rangle ,  \label{eq:CF7}
\end{equation}%
where $J_{\mu }^{\chi _{c1}}(x)$ is the interpolating current for the
axial-vector meson $\chi _{c1}(1P)$%
\begin{equation}
J_{\mu }^{\chi _{c1}}(x)=\overline{c}_{j}(x)\gamma _{5}\gamma _{\mu
}c_{j}(x).  \label{eq:Curr1}
\end{equation}

In terms of the physical parameters of involved particles this correlation
function has the form%
\begin{eqnarray}
&&\widetilde{\Pi }_{\mu }^{\mathrm{Phys}}(p,p^{\prime })=g_{5}(q^{2})\frac{%
\widetilde{f}\widetilde{m}f_{2}m_{2}^{2}f_{3}m_{3}}{2m_{c}\left( p^{2}-%
\widetilde{m}^{2}\right) \left( p^{\prime 2}-m_{3}^{2}\right) }\frac{1}{%
q^{2}-m_{2}^{2}}\left[ \frac{\widetilde{m}^{2}-m_{3}^{2}-q^{2}}{2m_{3}^{2}}%
p_{\mu }^{\prime }-q_{\mu }\right] +\cdots .  \notag \\
&&  \label{eq:CF8}
\end{eqnarray}%
In Eq.\ (\ref{eq:CF8}) $m_{3}$ and $f_{3}$ are the mass and decay constant
of the meson $\chi _{c1}(1P)$, respectively. To derive $\Pi _{\mu }^{\mathrm{%
Phys}}(p,p^{\prime })$, we have used the matrix elements of the molecule $%
\mathcal{M}_{2}$ and meson $\eta _{c}$, as well as new matrix elements
\begin{equation}
\langle 0|J_{\mu }^{\chi _{c1}}|\chi _{c1}(p^{\prime })\rangle
=f_{3}m_{3}\varepsilon _{\mu }^{\ast }(p^{\prime }),
\end{equation}%
and
\begin{equation}
\langle \eta _{c}(q)\chi _{c1}(p^{\prime })|\mathcal{M}_{2}(p)\rangle
=g_{5}(q^{2})p\cdot \varepsilon ^{\ast }(p^{\prime }),
\end{equation}%
where $\varepsilon _{\mu }^{\ast }(p^{\prime })$ is the polarization vector
of $\chi _{c1}(1P)$.

In terms of $c$-quark propagators the correlator $\widetilde{\Pi }_{\mu }^{%
\mathrm{OPE}}(p,p^{\prime })$ has the form Eq.\ (\ref{eq:QCD5}). The sum
rule for $g_{5}(q^{2})$ is derived using amplitudes corresponding to terms $%
\sim p_{\mu }^{\prime }$ in $\Pi _{\mu }^{\mathrm{Phys}}(p,p^{\prime })$ and
$\Pi _{\mu }^{\mathrm{OPE}}(p,p^{\prime })$.

In numerical analysis, the parameter $M_{2}^{2}$ and $s_{0}^{\prime }$ in
the $\chi _{c1}$ channel are chosen in the following way
\begin{equation}
M_{2}^{2}\in \lbrack 4,5]~\mathrm{GeV}^{2},\ s_{0}^{\prime }\in \lbrack
13,14]~\mathrm{GeV}^{2}.  \label{eq:Wind5}
\end{equation}%
For the parameters of the fit function $\mathcal{G}_{5}(Q^{2})$, we get $%
\mathcal{G}_{5}^{0}=6.02$, $c_{5}^{1}=3.16$, and $c_{5}^{2}=-3.88$. Then,
the strong coupling $g_{5}$ is equal to
\begin{equation}
g_{5}\equiv \mathcal{G}_{5}(-m_{2}^{2})=2.9\pm 0.6.
\end{equation}%
The width of the decay $\mathcal{M}_{2}\rightarrow \eta _{c}\chi _{c1}(P)$ \
can be calculated by means of the expression
\begin{equation}
\Gamma \left[ \mathcal{M}_{2}\rightarrow \eta _{c}\chi _{c1}(P)\right]
=g_{5}^{2}\frac{\lambda _{5}^{3}}{24\pi m_{3}^{2}},  \label{eq:DW3}
\end{equation}%
where $\lambda _{5}=\lambda (\widetilde{m},m_{3},m_{2})$. It is not
difficult to find that
\begin{equation}
\Gamma \left[ \mathcal{M}_{2}\rightarrow \eta _{c}\chi _{c1}(P)\right]
=(16\pm 5)~\mathrm{MeV}.  \label{eq:DW4}
\end{equation}

For studying the decay $\mathcal{M}_{2}\rightarrow \chi _{c0}\chi _{c0}$, we
consider the correlation function
\begin{equation}
\widetilde{\Pi }_{\chi _{c0}}(p,p^{\prime })=i^{2}\int
d^{4}xd^{4}ye^{ip^{\prime }y}e^{-ipx}\langle 0|\mathcal{T}\{J^{\chi
_{c0}}(y)J^{\chi _{c0}}(0)\widetilde{J}^{\dagger }(x)\}|0\rangle ,
\label{eq:CF9}
\end{equation}%
with $J^{\chi _{c0}}(x)$ being the interpolating current for the scalar
meson $\chi _{c0}$%
\begin{equation}
J^{\chi _{c0}}(x)=\overline{c}_{i}(x)c_{i}(x).
\end{equation}

The explicit expression of the correlator $\widetilde{\Pi }_{\chi _{c0}}^{%
\mathrm{OPE}}(p,p^{\prime })$ can be found in Eq.\ (\ref{eq:QCD6}). Remaning
operations are performed in the context of the standard approach. Thus, in
numerical computations, the parameters $M_{2}^{2}$ and $s_{0}^{\prime }$ in
the $\chi _{c0}$ channel are chosen in the form
\begin{equation}
M_{2}^{2}\in \lbrack 4,5]~\mathrm{GeV}^{2},\ s_{0}^{\prime }\in \lbrack
14,14.9]~\mathrm{GeV}^{2},
\end{equation}%
where $s_{0}^{\prime }$ is restricted by the mass of the charmonium $\chi
_{c0}(3860)$. The coupling $g_{6}$ that corresponds to the vertex $\mathcal{M%
}_{2}\chi _{c0}\chi _{c0}$ is extracted at $Q^{2}=-m_{4}^{2}$ of the fit
function $\mathcal{G}_{6}(Q^{2})$ with parameters $\mathcal{G}_{6}^{0}=0.63$%
, $c_{6}^{1}=2.83$, and $c_{6}^{2}=-3.03$.

The strong coupling $g_{6}$ is found equal to
\begin{equation}
g_{6}\equiv \mathcal{G}_{6}(-m_{4}^{2})=(2.7\pm 0.43)\times 10^{-1}\ \mathrm{%
GeV}^{-1}.
\end{equation}%
The partial width of the decay $\mathcal{M}_{2}\rightarrow \chi _{c0}\chi
_{c0}$ is calculated by means of the formula%
\begin{equation}
\Gamma \left[ \mathcal{M}_{2}\rightarrow \chi _{c0}\chi _{c0}\right]
=g_{6}^{2}\frac{m_{4}^{2}\lambda _{6}}{8\pi }\left( 1+\frac{\lambda _{6}^{2}%
}{m_{4}^{2}}\right) ,
\end{equation}%
where $\lambda _{6}=\lambda (\widetilde{m},m_{4},m_{4})$. Numerical analyses
yield%
\begin{equation}
\Gamma \left[ \mathcal{M}_{2}\rightarrow \chi _{c0}\chi _{c0}\right] =(22\pm
5)\times 10^{-1}~\mathrm{MeV}.
\end{equation}%
The partial widths of the six decays considered in this section are
collected in Table\ \ref{tab:Channels}.

Using these results, we estimate the full width of $\mathcal{M}_{2}$
\begin{equation}
\Gamma _{\mathcal{M}_{2}}=(138\pm 18)~\mathrm{MeV}.  \label{eq:FW2}
\end{equation}%
This prediction can be confronted with the data of the experimental groups.


\section{Discussion and concluding notes}

\label{sec:Disc}

In this article, we studied the hadronic molecules $\mathcal{M}_{1}=\eta
_{c}\eta _{c}$ and $\mathcal{M}_{2}=\chi _{c0}\chi _{c0}$ and calculated
their masses and full widths. The masses of these structures were extracted
from the QCD two-point sum rules. To evaluate full widths of $\mathcal{M}%
_{1} $ and $\mathcal{M}_{2}$, we applied the three-point sum rule method. We
analyzed two decay channels of the molecule $\mathcal{M}_{1}$. In the case
of $\mathcal{M}_{2}$ state, we took into account six kinematically allowed
decay modes of this molecule.

Our predictions for the mass $m=(6264\pm 50)~\mathrm{MeV}$ and width $\Gamma
_{\mathcal{M}_{1}}=(320\pm 72)~\mathrm{MeV}$ of the molecule $\mathcal{M}%
_{1} $ are consistent with the data of the ATLAS Collaboration which found
for these parameters
\begin{equation}
m^{\mathrm{ATL}}=6220\pm 50_{-50}^{+40}~\mathrm{MeV},\ \ \Gamma ^{\mathrm{ATL%
}}=310\pm 120_{-80}^{+70}~\mathrm{MeV}.
\end{equation}%
These results allow us to interpret the lowest resonance $X(6200)$ with
great confidence as the molecule $\eta _{c}\eta _{c}$.

The $\mathcal{M}_{2}=\chi _{c0}\chi _{c0}$ state has the mass and width%
\begin{equation}
\widetilde{m}=(6954\pm 50)~\mathrm{MeV},\ \Gamma _{\mathcal{M}_{2}}=(138\pm
18)~\mathrm{MeV}.
\end{equation}%
The mass $\widetilde{m}$ of the molecule $\mathcal{M}_{2}$ within errors of
computations agrees with the mass of the resonance $X(6900)$ measured by the
LHCb-ATLAS-CMS Collaborations, through the central value for $\widetilde{m}$
is a little over the relevant data. It is convenient to compare $\widetilde{m%
}$ and $\Gamma _{\mathcal{M}_{2}}$ with the CMS data
\begin{equation}
m^{\mathrm{CMS}}=(6927\pm 9\pm 4)~\mathrm{MeV},\ \Gamma ^{\mathrm{CMS}%
}=(122_{-21}^{+24}\pm 18)~\mathrm{MeV}.
\end{equation}%
One sees that the molecule $\mathcal{M}_{2}$ is a serious candidate to the
resonance $X(6900)$. The $X(6900)$ was also examined in our paper \cite%
{Agaev:2023gaq} in the context of the diquark-antidiquark model. The
predictions for the mass $m=(6928\pm 50)~\mathrm{MeV}$ and width $\widetilde{%
\Gamma }_{\mathrm{4c}}=(112\pm 21)~\mathrm{MeV}$ of the scalar tetraquark $%
T_{\mathrm{4c}}$ built of pseudoscalar constitutes are consistent with the
CMS data as well. These circumstances make a linear superposition of the
structures $\mathcal{M}_{2}$ and $T_{\mathrm{4c}}$ as one of the reliable
scenarios for the resonance $X(6900)$.

\appendix*

\section{ Heavy-quark propagator $S_{Q}^{ab}(x)$ and correlation functions}

\renewcommand{\theequation}{\Alph{section}.\arabic{equation}} \label{sec:App}

In the present study, for the heavy quark propagator $S_{Q}^{ab}(x)$ ($Q=c,\
b$), we use
\begin{eqnarray}
&&S_{Q}^{ab}(x)=i\int \frac{d^{4}k}{(2\pi )^{4}}e^{-ikx}\Bigg \{\frac{\delta
_{ab}\left( {\slashed k}+m_{Q}\right) }{k^{2}-m_{Q}^{2}}-\frac{%
g_{s}G_{ab}^{\alpha \beta }}{4}\frac{\sigma _{\alpha \beta }\left( {\slashed %
k}+m_{Q}\right) +\left( {\slashed k}+m_{Q}\right) \sigma _{\alpha \beta }}{%
(k^{2}-m_{Q}^{2})^{2}}  \notag \\
&&+\frac{g_{s}^{2}G^{2}}{12}\delta _{ab}m_{Q}\frac{k^{2}+m_{Q}{\slashed k}}{%
(k^{2}-m_{Q}^{2})^{4}}+\cdots \Bigg \}.
\end{eqnarray}%
Here, we have used the notations
\begin{equation}
G_{ab}^{\alpha \beta }\equiv G_{A}^{\alpha \beta }\lambda _{ab}^{A}/2,\ \
G^{2}=G_{\alpha \beta }^{A}G_{A}^{\alpha \beta },\
\end{equation}%
where $G_{A}^{\alpha \beta }$ is the gluon field-strength tensor, and $%
\lambda ^{A}$ are the Gell-Mann matrices. The indices $A,B,C$ run in the
range $1,2,\ldots 8$.

The correlation function $\widetilde{\Pi }^{\mathrm{OPE}}(p)$ used to
calculate the mass and current coupling of the molecule $\mathcal{M}_{2}$:%
\begin{eqnarray}
&&\widetilde{\Pi }^{\mathrm{OPE}}(p)=i\int d^{4}xe^{ipx}\left\{ \mathrm{Tr}%
\left[ S_{c}^{ba^{\prime }}(x)S_{c}^{a^{\prime }b}(-x)\right] \mathrm{Tr}%
\left[ S_{c}^{ab^{\prime }}(x)S_{c}^{b^{\prime }a}(-x)\right] -\mathrm{Tr}%
\left[ S_{c}^{bb^{\prime }}(x)S_{c}^{b^{\prime }a}(-x)\right. \right.  \notag
\\
&&\left. \times S_{c}^{aa^{\prime }}(x)S_{c}^{a^{\prime }b}(-x)\right] -%
\mathrm{Tr}\left[ S_{c}^{ba^{\prime }}(x)S_{c}^{a^{\prime
}a}(-x)S_{c}^{ab^{\prime }}(x)S_{c}^{b^{\prime }b}(-x)\right] +\mathrm{Tr}%
\left[ S_{c}^{bb^{\prime }}(x)S_{c}^{b^{\prime }b}(-x)\right]  \notag \\
&&\left. \times \mathrm{Tr}\left[ S_{c}^{aa^{\prime }}(x)S_{c}^{a^{\prime
}a}(-x)\right] \right\} .  \label{eq:QCD2}
\end{eqnarray}%
The correlators $\widetilde{\Pi }_{\mu \nu }^{\mathrm{Phys}}(p,p^{\prime })$
and $\widetilde{\Pi }_{\mu \nu }^{\mathrm{OPE}}(p,p^{\prime })$ necessary to
explore the decays $\mathcal{M}_{2}\rightarrow J/\psi J/\psi (\psi ^{\prime
})$:%
\begin{eqnarray}
&&\widetilde{\Pi }_{\mu \nu }^{\mathrm{Phys}}(p,p^{\prime })=\frac{\langle
0|J_{\mu }^{\psi }|J/\psi (p^{\prime })\rangle }{p^{\prime 2}-m_{1}^{2}}%
\frac{\langle 0|J_{\nu }^{\psi }|J/\psi (q)\rangle }{q^{2}-m_{1}^{2}}\langle
J/\psi (p^{\prime })J/\psi (q)|\mathcal{M}_{2}(p)\rangle \frac{\langle
\mathcal{M}_{2}(p)|\widetilde{J}^{\dagger }|0\rangle }{p^{2}-\widetilde{m}%
^{2}}  \notag \\
&&+\frac{\langle 0|J_{\mu }^{\psi }|\psi (p^{\prime })\rangle }{p^{\prime
2}-m_{1}^{\ast 2}}\frac{\langle 0|J_{\nu }^{\psi }|J/\psi (q)\rangle }{%
q^{2}-m_{1}^{2}}\langle \psi (p^{\prime })J/\psi (q)|\mathcal{M}%
_{2}(p)\rangle \frac{\langle \mathcal{M}_{2}(p)|\widetilde{J}^{\dagger
}|0\rangle }{p^{2}-\widetilde{m}^{2}}+\cdots ,  \label{eq:CF3A}
\end{eqnarray}%
and%
\begin{eqnarray}
&&\widetilde{\Pi }_{\mu \nu }^{\mathrm{OPE}}(p,p^{\prime })=2i^{2}\int
d^{4}xd^{4}ye^{ip^{\prime }y}e^{-ipx}\left\{ \mathrm{Tr}\left[ \gamma _{\mu
}S_{c}^{ia}(y-x)S_{c}^{ai}(x-y)\right] \mathrm{Tr}\left[ \gamma _{\nu
}S_{c}^{jb}(-x)S_{c}^{bj}(x){}\right] \right.  \notag \\
&&\left. -\mathrm{Tr}\left[ \gamma _{\mu
}S_{c}^{ia}(y-x)S_{c}^{aj}(x){}\gamma _{\nu }S_{c}^{jb}(-x)S_{c}^{bi}(x-y)%
\right] \right\} .  \label{eq:QCD3}
\end{eqnarray}%
The correlation functions $\widetilde{\Pi }^{\mathrm{Phys}}(p,p^{\prime })$
and $\widetilde{\Pi }^{\mathrm{OPE}}(p,p^{\prime })$ used in the analysis of
the decays $\mathcal{M}_{2}\rightarrow \eta _{c}\eta _{c}(\eta _{c}(2S))$
\begin{eqnarray}
&&\widetilde{\Pi }^{\mathrm{Phys}}(p,p^{\prime })=\frac{\langle 0|J^{\eta
_{c}}|\eta _{c}(p^{\prime })\rangle }{p^{\prime 2}-m_{2}^{2}}\frac{\langle
0|J^{\eta _{c}}|\eta _{c}(q)\rangle }{q^{2}-m_{2}^{2}}\langle \eta
_{c}(p^{\prime })\eta _{c}(q)|\mathcal{M}_{2}(p)\rangle \frac{\langle
\mathcal{M}_{2}(p)|J^{\dagger }|0\rangle }{p^{2}-\widetilde{m}^{2}}  \notag
\\
&&+\frac{\langle 0|J^{\eta _{c}}|\eta _{c}(2S)(p^{\prime })\rangle }{%
p^{\prime 2}-m_{2}^{\ast 2}}\frac{\langle 0|J^{\eta _{c}}|\eta
_{c}(q)\rangle }{q^{2}-m_{2}^{2}}\langle \eta _{c}(2S)(p^{\prime })\eta
_{c}(q)|\mathcal{M}_{2}(p)\rangle \frac{\langle \mathcal{M}%
_{2}(p)|J^{\dagger }|0\rangle }{p^{2}-\widetilde{m}^{2}}+\cdots ,
\label{eq:CF5A}
\end{eqnarray}%
and
\begin{equation}
\widetilde{\Pi }^{\mathrm{OPE}}(p,p^{\prime })=-2\int
d^{4}xd^{4}ye^{ip^{\prime }y}e^{-ipx}\mathrm{Tr}\left[ \gamma
_{5}S_{c}^{ia}(y-x)S_{c}^{aj}(x){}\gamma _{5}S_{c}^{jb}(-x)S_{c}^{bi}(x-y)%
\right] .  \label{eq:QCD4}
\end{equation}

The correlation function $\widetilde{\Pi }_{\mu }^{\mathrm{OPE}}(p,p^{\prime
})$ for the process $\mathcal{M}_{2}\rightarrow \eta _{c}\chi _{1c}(1P)$ is
given by the formula%
\begin{equation}
\widetilde{\Pi }_{\mu }^{\mathrm{OPE}}(p,p^{\prime })=-2i^{3}\int
d^{4}xd^{4}ye^{ip^{\prime }y}e^{-ipx}\mathrm{Tr}\left[ \gamma _{\mu }\gamma
_{5}S_{c}^{ia}(y-x)S_{c}^{aj}(x){}\gamma _{5}S_{c}^{jb}(-x)S_{c}^{bi}(x-y)%
\right] .  \label{eq:QCD5}
\end{equation}

The function $\widetilde{\Pi }_{\chi _{c0}}^{\mathrm{OPE}}(p,p^{\prime })$
for the decay $\mathcal{M}_{2}\rightarrow \chi _{c0}\chi _{c0}$ is:%
\begin{eqnarray}
\widetilde{\Pi }_{\chi _{c0}}^{\mathrm{OPE}}(p,p^{\prime }) &=&2i^{2}\int
d^{4}xd^{4}ye^{ip^{\prime }y}e^{-ipx}\left\{ \mathrm{Tr}\left[
S_{c}^{ia}(y-x)S_{c}^{ai}(x-y)\right] \mathrm{Tr}\left[
S_{c}^{jb}(-x)S_{c}^{bj}(x)\right] \right.  \notag \\
&&\left. -\mathrm{Tr}\left[
S_{c}^{ia}(y-x)S_{c}^{aj}(x)S_{c}^{jb}(-x)S_{c}^{bi}(x-y)\right] \right\}.
\label{eq:QCD6}
\end{eqnarray}

\textbf{Data Availability Statement:} No Data associated in the manuscript.

\end{document}